\documentclass[aps,twocolumn]{revtex4}

\usepackage{graphicx}
\usepackage{dcolumn}
\usepackage{bm}
\usepackage{amsmath}
\begin{document}

\newcommand{\bk}{{\bf k}}
\newcommand{\bp}{{\bf p}}
\newcommand{\bv}{{\bf v}}
\newcommand{\bq}{{\bf q}}
\newcommand{\bQ}{{\bf Q}}
\newcommand{\br}{{\bf r}}
\newcommand{\bR}{{\bf R}}
\newcommand{\bB}{{\bf B}}
\newcommand{\bA}{{\bf A}}
\newcommand{\bK}{{\bf K}}
\newcommand{\vd}{{v_\Delta}}
\newcommand{\tr}{{\rm Tr}}
\newcommand{\kslash}{\not\!k}
\newcommand{\qslash}{\not\!q}
\newcommand{\pslash}{\not\!p}
\newcommand{\rslash}{\not\!r}
\newcommand{\bs}{{\bar\sigma}}

\title{Magnetic field induced charge and spin instabilities in 
cuprate superconductors}

\author{M. Franz\rlap,$^1$ D. E. Sheehy$^1$ and Z. Te\v{s}anovi\'c$^2$}
\affiliation{$^1$Department of Physics and Astronomy,
University of British Columbia, Vancouver, BC, Canada V6T 1Z1\\
$^2$Department of Physics and Astronomy,
Johns Hopkins University, Baltimore, MD 21218}

\date{\today}

\begin{abstract}
A $d$-wave superconductor, subject to strong phase fluctuations, is known
to suffer an  antiferromagnetic instability closely related to the chiral 
symmetry
breaking in (2+1)-dimensional quantum electrodynamics (QED$_3$). Based
on this idea we formulate a ``QED$_3$ in a box'' theory of
{\em local} instabilities of a $d$-wave superconductor in the 
vicinity of a single pinned vortex undergoing quantum fluctuations. 
As a generic outcome we find an incommensurate 2D spin
density wave forming in the neighborhood of a vortex with a concomitant 
``checkerboard'' pattern in the local electronic density of states, in 
agreement with recent neutron scattering and tunneling spectroscopy 
measurements.
\end{abstract}

\maketitle

Among the open questions in modern condensed matter physics, few have 
inspired more theoretical effort than the emergence of a superconducting
state from the doped antiferromagnetic (AF) insulator~\cite{anderson1}.
Recently,  using an ``inverted'' approach to the problem \cite{balents1,ft1},
it has been shown that AF order arises naturally when the superconducting 
order in a $d$-wave superconductor ($d$SC) is destroyed by vortex-antivortex
fluctuations \cite{herbut1,ft2}.
As we shall discuss, the implications of these theories transcend the 
possibility of providing a
route to understanding the destruction of superconductivity in 
strongly-underdoped cuprates; indeed, they also apply to 
the problem of {\em local} field-induced vortices within the superconducting 
state.

Recent neutron scattering~\cite{lake1,lake2} and scanning tunneling 
spectroscopy (STS)~\cite{davis1} 
experiments have revealed the presence of local AF and charge 
order in the vicinity of field-induced vortices.
Existing theoretical treatments~\cite{arovas1,demler1,ting1,zhang1,demler2}  
of vortex-induced AF ordering rely on the proximity of the system to a 
quantum critical point.  Within such treatments, it is the suppression of the 
SC order parameter near the vortex 
cores that leads to the nucleation of islands of AF order.
Here we  present an alternative scenario in 
which the AF order is brought about by {\em local quantum 
fluctuations} of a vortex around its equilibrium position.
In the present theory
there is no competition between the SC amplitude and AF order: 
the latter arises purely from the presence of vortex fluctuations and is a 
genuine low-energy phenomenon taking place on lengthscales much longer than
the core size.

It is a well-known fact that the low superfluid density in cuprates
makes the SC order 
vulnerable to phase fluctuations~\cite{emery1,corson1,ong1}.
This observation has inspired theories in which the pseudogap state is modeled 
as a phase-disordered $d$-wave 
superconductor~\cite{balents1,ft1,franz1,kwon1}, such that the 
demise of superconductivity is brought about by the unbinding and 
proliferation of the topological defects -- vortices -- in the phase of the 
SC order parameter. 
It has been pointed 
out~\cite{ft1} that fluctuating vortices produce 
a non-trivial Berry-phase interaction between the quasiparticles of the 
underlying $d$SC.
This interaction is described
in terms of a massless non-compact gauge field $a_\mu$, minimally coupled to 
the Dirac fermions representing the low-energy quasiparticle
 excitations of the 
system.  Within the theory of Ref.~\cite{ft1} which maps the problem
onto (2+1)-dimensional quantum electrodynamics (QED$_3$), it is this 
interaction that destroys the Fermi liquid nature of quasiparticles in the 
pseudogap state and ultimately drives the AF instability~\cite{herbut1,ft2}. 
Remarkably, both the `algebraic' Fermi liquid describing the symmetric 
pseudogap phase and the 
antiferromagnet emerge from the same QED$_3$ theory \cite{ft1}.

Here, we use the philosophy and formalism developed in 
Refs.~\cite{ft1,herbut1,ft2} to model quasiparticle excitations in the {\em 
superconducting state} in the spatial region close to a single field-induced 
vortex undergoing fluctuations around its equilibrium position. 
We call this model ``QED$_3$ in a box''.
We note that there
exists direct experimental evidence that individual vortices indeed undergo
significant quantum fluctuations~\cite{hoogenboom1}. We find that, under
generic conditions, interactions generated by such fluctuating vortex
lead to {\em local} instability of the superconducting state which takes
form of a 2D incommensurate AF spin density wave (SDW) with a wave vector 
tied to the positions of the nodes in the underlying $d$-wave gap. 

In order to motivate our model for a single vortex we first review the 
treatment of the AF instability in QED$_3$ and  reformulate
it in a way that will be more suitable for our present purposes. We start from 
the Euclidean  QED$_3$ action $S=\int d^3x{\cal L}_D$ with
\begin{equation}
{\cal L}_D\equiv \sum_{l=1}^{N} 
\bar\Psi_l(x) \gamma_\mu (i\partial_\mu -a_\mu)\Psi_l(x) + {\cal L}_B[a(x)],
\label{l1}
\end{equation}
describing the low-energy fermionic excitations of a $d$-wave SC coupled 
to fluctuating vortices represented by the gauge field $a_\mu$~\cite{ft1}.
Here, $\Psi_l(x)$ is a four component Dirac spinor representing the fermionic 
excitations associated with a pair of antipodal
nodes, $x=(\tau,{\br})$ denotes the space-time coordinate, and 
$\gamma_\mu$ ($\mu=0,1,2$) are the gamma matrices satisfying
$\{\gamma_\mu,\gamma_\nu\}=2\delta_{\mu\nu}$.  The number  $N$ of 
fermion species is equal to $2$ for single-layer cuprates; $N=4,6,...$
for bilayer, trilayer and multilayer materials.
The Lagrangian ${\cal L}_B$ encodes the dynamics
of the gauge field $a_\mu$ and is given by 
${\cal L}_B[a]=\Pi_{\mu\nu}(q)a_\mu(q)a_\nu(-q)$ with
\begin{equation}
\Pi_{\mu\nu}(q)=\left(m_a+ \frac{N}{8}|q|\right)
\left(\delta_{\mu\nu}-\frac{q_\mu q_\nu}{q^2}\right).
\label{ber}
\end{equation}
The gauge field mass $m_a$ vanishes when  vortices
are unbound (i.e., in the pseudogap regime or, in the present situation, near 
a single fluctuating vortex)
and is finite in the superconducting state where vortices appear
only in tightly bound loops or pairs.

In the standard treatment~\cite{herbut1,ft2} the AF order occurs via the 
phenomenon of chiral symmetry breaking~\cite{pisarski1,appelquist1,nash1}
 in the QED$_3$ Lagrangian (\ref{l1}). The instability is signaled by the
spontaneous
generation of fermion mass, $m_D$, which is interpreted in our context as
the onset of SDW gap~\cite{herbut1,ft2} for the original Bogoliubov 
quasiparticles. 
The most general, nonperturbative treatment of mass-generation in QED$_3$ 
obtains $m_D$ 
as a solution of a self-consistent Dyson-Schwinger equation. Here we shall 
follow a slightly simpler route which leads to the same result and has the advantage
of being more easily generalizable to the present problem.  In 
Eq.~(\ref{l1}) we integrate out the gauge field to
obtain the  following fermionic effective action:
\begin{eqnarray}
S_{\rm eff}&=&\int d^3x 
\bar\Psi(x) \gamma_\mu i\partial_\mu\Psi(x) \label{l2} \\
&-& \int d^3x\int d^3y J_\mu(x)D_{\mu\nu}(x-y)J_\nu(y),
\nonumber
\end{eqnarray}
where $J_\mu(x)\equiv\bar\Psi(x)\gamma_\mu\Psi(x)$ is the fermion 3-current
and $D_{\mu\nu}(x)$ is the Fourier transform of the gauge boson propagator
$D_{\mu\nu}(q)=\Pi_{\mu\nu}^{-1}(q)$. Henceforth we shall focus on a
single pair of nodes and thus drop the  nodal index $l$.
The integrand of the 
interaction term may be written as $D_{\mu\nu}(x-y)\tr[\Psi(y)\bar\Psi(x)
\gamma_\mu\Psi(x)\bar\Psi(y)\gamma_\nu]$ where the trace is taken over the 
spinor indices. This form suggests a Hartree-Fock (HF) approach in which we 
decouple the 4-fermion interaction to obtain 
$D_{\mu\nu}(x-y)\tr[\Psi(y)\bar\Psi(x)\gamma_\mu G_0(x,y)\gamma_\nu]$ with 
$G_0(x,y)=\langle\Psi(x)\bar\Psi(y)\rangle$ and the average is taken
with respect to the HF effective action to be specified shortly. To make the
structure of the interaction term more transparent we 
 utilize the relative and center of
mass coordinates $r=x-y$ and $R=(x+y)/2$ to write it as
\begin{equation}
\int d^3R\,\int d^3r\,\tr\left[
\Psi(R_+)\bar\Psi(R_-)\gamma_\mu G_0(R,r)\gamma_\nu\right]D_{\mu\nu}(r),
\label{int1}
\end{equation}
where $R_\pm\equiv R\pm\frac{r}{2}$. In the {\em uniform} system the
Green's function is independent of $R$, $G_0(R,r)=G_0(r)$. Furthermore,
both $G_0(r)$ and $D_{\mu\nu}(r)$ are strongly peaked at $r\to 0$. The 
dominant contribution to the interaction therefore comes from this region and
we may write (\ref{int1}) as~\cite{remark1}
\begin{equation}
\int d^3R\bar\Psi(R)\Psi(R)\int d^3r
\gamma_\mu G_0(r)\gamma_\nu D_{\mu\nu}(r).
\label{int2}
\end{equation}
We have dropped the trace since the interaction is proportional to the unit matrix
in the spinor space. 

Inspection of Eq.~(\ref{int2}) suggests the following HF effective action
and self-consistency condition: 
\begin{subequations}
\begin{eqnarray}
S_{\rm HF}&=&\int d^3x 
\bar\Psi(x)( \gamma_\mu i\partial_\mu-im_D)\Psi(x),
\label{l3}
\\
im_D&=&\frac{1}{4}\tr\int d^3r
\gamma_\mu G_0(r)\gamma_\nu D_{\mu\nu}(r).
\label{self}
\end{eqnarray}
\end{subequations}
The last integral is easily evaluated by going to momentum space and Eq.\ 
(\ref{self}) becomes $m_D=(8m_D/N\pi^2)\ln(\Lambda/m_D)$, where 
$\Lambda$ is the high-momentum cutoff. This yields a nontrivial solution
\begin{equation}
m_D=\Lambda e^{-N\pi^2/8},
\label{mass}
\end{equation}
in agreement with the classic result of Pisarski~\cite{pisarski1}. 
More sophisticated treatments
\cite{appelquist1,nash1} based on the Schwinger-Dyson equation give a finite critical value of $N_c$ above which 
no mass is generated; however, for our purposes the level of approximation embodied by Eq.~(\ref{mass}) 
will be sufficient.

We have thus seen that, in a uniform system, fluctuating vortices lead to the formation
of SDW order.
The challenge we now face is twofold: (i) we must adapt the above treatment 
to the case of a single
fluctuating vortex, and (ii) since we seek to study the commensuration effects
present in real materials, we must formulate the corresponding theory on the 
lattice. To address (i) let us denote by $\ell_v$ the characteristic 
length scale over which the vortex fluctuates around its classical 
equilibrium position. Within this length scale, the Berry-phase interaction 
between quasiparticles (and hence tendency towards SDW ordering) will be strong.
 We model this by taking in this region the gauge field  
to be {\em massless}. On the other hand at distances well beyond $\ell_v$, 
quasiparticles feel no interaction and we model this by gauge field having a 
large mass $m_a$. In particular we take, 
\begin{equation}
m_a(\bR)=\Delta_0\left(\frac{|\bR|}{\ell_v}\right)^n,
\label{gaugemass}
\end{equation}
where $\Delta_0$ is an energy scale which we take to be 
the maximum superconducting gap, $|\bR|$ is the 
distance from the vortex equilibrium position and $n$ is a positive exponent.
(We use $n=2$ but our numerical calculations below are largely insensitive
to the exact value of $n$.)

To address (ii), (i.e.,~to put the theory on the lattice) we recall that the 
effective action (\ref{l1}) and its HF version Eq.~(\ref{l3}) descend from a 
model of a lattice 
$d$SC linearized near the nodes of the gap. We therefore consider the
corresponding lattice Hamiltonian enriched by the ``mass'' term present 
in Eq.~(\ref{l3}), to represent the HF decoupled Berry phase interaction.  Thus we have,
\begin{equation}
H_{\rm HF}=\sum_\sigma\sum_{\langle ij\rangle}
\Phi^\dagger_{i\sigma}{\cal H}_{ij}^\sigma \Phi_{j\sigma}.
\label{h1}
\end{equation}
Here $\Phi^\dagger_{i\sigma}=(c^\dag_{i\sigma},c_{i\bs})$, $c^\dag_{i\sigma}$
represents the electron creation operator at lattice site $i$, 
spin index $\bs=-\sigma$, and  
\begin{equation}
{\cal H}_{ij}^\sigma=\left(
\begin{array}{cc}
-t_{ij}+\delta_{ij}(m_{i\sigma}-\mu) & \Delta_{ij} \\
\Delta^*_{ij} & t_{ij}-\delta_{ij}(m_{i\bs}-\mu)
\end{array}
\right), \nonumber
\end{equation}
with $t_{ij}$ the tight binding hopping amplitude, $\Delta_{ij}$ the
SC gap, $\mu$ chemical potential, and $m_{i\sigma}$ the local spin 
magnetization representing the mass gap $m_D$ in Eq.\ (\ref{l3}).

We diagonalize $H_{\rm HF}$ by the generalized Bogoliubov 
transformation $c_{i\sigma}=\sum_n[u_{n\sigma}(\br_i)\gamma_{n\sigma}
+\sigma v^*_{n\bs}(\br_i)\gamma^\dag_{n\bs}]$, where 
$\chi_{n\sigma}(\br_i)\equiv [u_{n\sigma}(\br_i),\sigma v_{n\sigma}(\br_i)]^T$
satisfy
\begin{equation}
\sum_j{\cal H}_{ij}^\sigma \chi_{n\sigma}(\br_j)=\epsilon_{n\sigma}
\chi_{n\sigma}(\br_i).
\label{eigen1}
\end{equation}
In terms of the $\chi_{n\sigma}$,
the self-consistency condition (\ref{self}) can be written as
\begin{eqnarray}
m_{i\sigma}&=&\sum_{n\sigma j}\sigma f(\epsilon_{n\sigma})
V_i(\br_j)u^*_{n\sigma}(\bR_i+\br_j)u_{n\sigma}(\bR_i-\br_j),
\nonumber
\\
V_i(\br)&\equiv&\frac{1}{4}\int_0^\infty d\tau e^{-\tau\epsilon_{n\sigma}}
\tr\left[\gamma_\mu D_{\mu\nu}(\tau,\br)\gamma_\nu\right].
\label{self2}
\end{eqnarray}
For $m_a\neq 0$ the above integral cannot be evaluated in  closed form. 
However, we find that it can be accurately approximated by a simple 
interpolation formula
\begin{equation}
V_i(\br)\simeq{\cal V}_0
\frac{c_1}
{r(rm_a(\bR_i)+c_1)(r\epsilon_{n\sigma}+c_1)}
\label{pot2}
\end{equation}
where $r=|\br|$, $c_1=2/\pi$ and ${\cal V}_0=16/N\pi^2$ for the case 
of an isotropic Dirac
cone ($t=\Delta$). In the physical case $t>\Delta$ the constant ${\cal V}_0$ 
will be modified somewhat and in what follows we treat it as an adjustable
parameter of the model measuring the strength of the interaction. It is 
interesting to note that, as seen from Eq.~(\ref{pot2}),
 in (2+1)D a gauge field mass does {\em not} lead to 
exponentially decaying interactions on  long length scales. 

To capture the effect of vortex fluctuations on the local superconducting order
we solve the eigenproblem~(\ref{eigen1}) numerically on a lattice of 
$M\times M$ sites and iterate to self-consistency using Eq.~(\ref{self2}).
For simplicity we consider only nearest-neighbor hopping amplitudes 
$t_{ij}=t$ and a
uniform $d$-wave gap $\Delta_{ij}=\pm\Delta_0$, with $+$ and 
$-$ signs referring to vertical and horizontal bonds respectively. 
We emphasize that the vortex, fluctuating around the equilibrium position at the center of our 
lattice, enters through the
position dependent gauge-field mass $m_a(\bR)$ given by Eq.\ 
(\ref{gaugemass}), which in turn enters the potential $V_i(\br)$ 
given by Eq.\ (\ref{pot2}). In the spirit of our working philosophy 
that the SDW 
order arises from the vortex fluctuations (and therefore from the gauge field),
we neglect at this stage any effects of the superflow around the vortex or 
suppression of $\Delta_{ij}$ in the core. Such effects are well understood 
and will be included in a future publication. We also neglect the effects
of changes in the fermionic spectrum due to the onset of SDW on 
the interaction mediated by Berry gauge field Eq.\ (\ref{pot2}).

\begin{figure}[t]
\includegraphics[width=8.5cm]{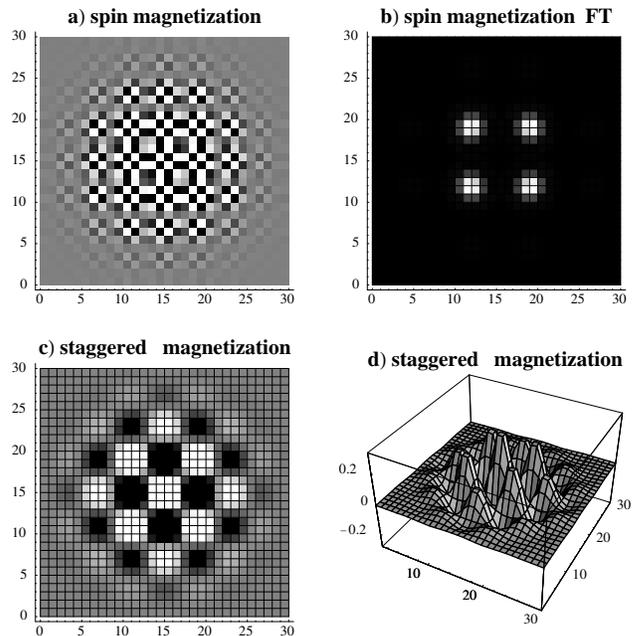}
\caption{\label{fig1}
Magnetization $M_i$ and staggered magnetization $M^S_i$ induced in the 
vicinity of the fluctuating vortex. Parameters used: $t=1$, $\Delta_0=0.7$,
$\mu=-1.6$ resulting in maximum gap of 2.3 and 
average charge density $n=0.62$ electrons per site,
${\cal V}_0=1.0$ and $\ell_v=12$; periodic boundary conditions.
}
\end{figure}

\begin{figure}[t]
\includegraphics[width=8.5cm]{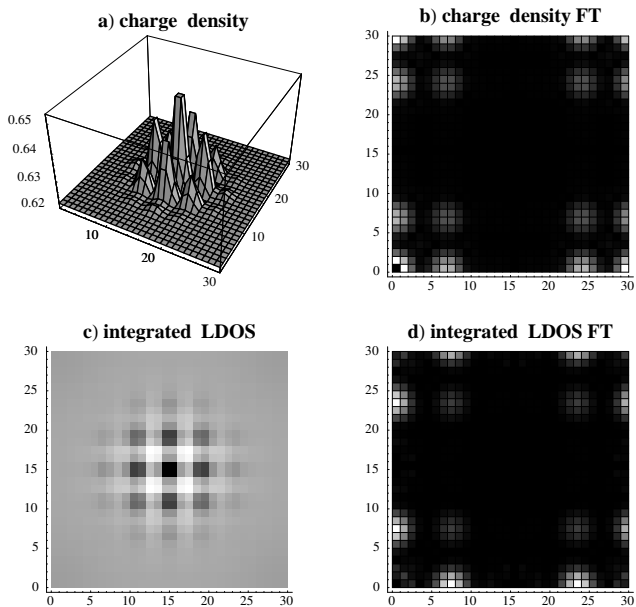}
\caption{\label{fig2}
Local charge density $n_i$ and integrated LDOS $S_{0.2}^{0.7}(i)$ for the same 
parameters as Fig. \ref{fig1}.
}
\end{figure}

The diagonalizations are performed using standard {\sc LAPACK} routines, which 
allow us to handle systems up to $40\times 40$ sites. Typically, 10-15 
iterations are
needed to ensure self-consistency in $m_{i\sigma}$. We use both periodic 
and free boundary conditions and find that they have negligible effect on 
the results reported below. Our typical results are summarized in Figs.~\ref{fig1} 
and \ref{fig2}, showing the spatial distributions of the spin 
magnetization $M_i=\sum_\sigma\sigma\langle c^\dag_{i\sigma}c_{i\sigma}
\rangle$, staggered spin magnetization 
$M^S_i=M_i(-1)^{x_i+y_i}$, local electron charge density $n_i=\sum_\sigma
\langle c^\dag_{i\sigma}c_{i\sigma}\rangle$, and energy integrated 
local density of states (LDOS) $S_{E_1}^{E_2}(i)=\int_{E_1}^{E_2}\rho_i(E)dE$
where $\rho_i(E)$ is the LDOS at site $i$, as well as their respective Fourier 
transforms (FTs).

Panel (a) in Fig. \ref{fig1} illustrates the ``2D'' incommensurate SDW pattern
emerging in the vicinity of a fluctuating vortex
with an $8\times 8$ unit cell containing islands of AF order separated 
by anti-phase domain walls [apparent in panel (c)]. The FT displayed in panel 
(b) reveals that this pattern can be thought of as a superposition of four 
1D SDWs with wave vectors $\bQ_{\rm SDW}=\pi(1\pm\delta_{\rm SDW},
1\pm\delta_{\rm SDW})$, $\delta_{\rm SDW}=\frac{1}{4}$. 
The size of $\delta_{\rm SDW}$ is doping dependent: it shrinks with 
increasing $\mu$ 
and vanishes at half filling $(\mu=1)$, giving rise to perfectly commensurate
AF SDW. We also find that for $\mu<-1.7$ the SDW becomes very weak for 
reasonable values of coupling ${\cal V}_0$: overdoped samples are less
susceptible to AF instability.

According to general symmetry arguments, a  spatial
modulation in the spin density generates a modulation in the charge density,
$\delta n_i\propto M_i^2$. For our 2D SDW pattern this implies that the
corresponding CDW will have a unit cell with half the area, rotated by 
$45^\circ$ relative to the unit cell of $M^S_i$. Indeed, panels 
(a) and (b) of Fig. \ref{fig2} confirm this general
expectation, showing a ``checkerboard'' CDW at principal wavevectors 
$\bQ_{\rm CDW}=\pi(\pm\delta_{\rm CDW},0),\pi(0,\pm\delta_{\rm CDW})$ with 
$\delta_{\rm CDW}=\frac{1}{2}$. A similar checkerboard pattern arises in the
integrated LDOS and is displayed in panels (c) and (d).

Our findings of a checkerboard pattern in LDOS are consistent with the recent 
STS experiments performed on Bi$_2$Sr$_2$CaCu$_2$O$_{8+\delta}$ (BSCCO)
crystals~\cite{davis1}. Our prediction is that the period of the pattern should
increase with underdoping and that the effect should vanish in the overdoped 
samples. Also, if the observed LDOS pattern is associated with electron 
density modulation in a single CuO layer, we predict that the corresponding 
neutron scattering peaks should be found at wavectors $\bQ_{\rm SDW}=
\pi(1\pm\frac{1}{4},1\pm\frac{1}{4})$. We note that neutron experiments
\cite{lake1,lake2} on
La$_{1.84}$Sr$_{0.16}$CuO$_4$ (LSCO) show peaks at different $k$-space 
positions, $\pi(1\pm\frac{1}{4},1)$ and $\pi(1,1\pm\frac{1}{4})$. Although the
findings of STS and neutron experiments are generally cited as being mutually 
consistent our analysis above indicates that this is not necessarily so:
for a genuine 2D SDW 
illustrated in Fig.~\ref{fig1} determination of the corresponding CDW
must consider the interference terms which cause the apparent $45^\circ$
rotation of the latter. In the absence of neutron measurements on BSCCO we see
two possible resolutions of this difficulty. First, it may be that the
CDW pattern is truly 2D and neutron scattering on BSCCO would find a pattern 
illustrated in Fig. \ref{fig1}(b). Second, it could be that STS sees an
incoherent superposition of two orthogonal 1D CDWs originating in two CuO 
layers comprising the BSCCO bilayer. This would explain the $x-y$ anisotropy 
reported in Ref.~\cite{davis1} and the neutron pattern would be consistent
with that observed in LSCO. Within our simple model such 1D solutions have
slightly higher energy than the 2D solutions reported above but it is possible 
that in a more complete model (using e.g.~a more realistic band structure) the
situation will be reversed.

To summarize, we have presented a ``QED$_3$ in a box'' theory for field 
induced spin and charge 
instabilities in cuprates, driven by local fluctuations of a pinned vortex.
Without any need to fine-tune parameters we find LDOS patterns in detailed
agreement with the tunneling data on BSCCO~\cite{davis1} and we relate
them in a plausible way to existing neutron experiments~\cite{lake1,lake2}. 
Other models~\cite{arovas1,demler1,ting1,zhang1,demler2} rely on the local 
suppression of the superconducting order parameter and would therefore predict similar
instabilities in the vicinity of impurities, grain boundaries, and sample 
edges where thus far no such effects have been observed. 

This research was supported in part by NSERC (MF,DES) and 
NSF Grant DMR00-94981 (ZT).


\begin{thebibliography}{10}
\bibitem{remark1} This approximation is equivalent to taking the
$k$-independent part of $\Sigma(k)$ in the conventional
treatment using the Dyson-Schwinger equation \cite{pisarski1}.
\end{thebibliography}
\end{document}